\documentclass[preprint,aps,showpacs,nofootinbib,preprintnumbers,amsmath,amssymb,superscriptaddress]{revtex4-2}

\usepackage{graphicx}
\usepackage{bm}
\usepackage{amssymb}
\usepackage{amsmath}
\usepackage{epsfig}
\usepackage{color}
\usepackage{mathtools}
\usepackage{cases}
\usepackage{booktabs}

\usepackage{dcolumn}
\usepackage{bm}
\usepackage{braket}

\usepackage[
colorlinks=true,
filecolor=black,
anchorcolor=blue,
linkcolor=blue,
citecolor=blue, 
urlcolor=blue,
linktocpage=true,
plainpages=false,
breaklinks=true,
pdfstartview=FitH
]{hyperref}

\begin{document}
\title{Gravitational Wave Birefringence in Symmetron Cosmology}
\author{Ze-Xuan Xiong\footnote{xiongzexuan22@mails.ucas.ac.cn}}
\affiliation{School of Fundamental Physics and Mathematical Sciences,  Hangzhou   Institute for Advanced Study, UCAS, Hangzhou 310024, China}
\affiliation{University of Chinese Academy of Sciences, 100190 Beijing, China}

\author{Da~Huang\footnote{Corresponding author: dahuang@bao.ac.cn}}
\affiliation{National Astronomical Observatories, Chinese Academy of Sciences, Beijing, 100101, China}
\affiliation{School of Fundamental Physics and Mathematical Sciences,  Hangzhou   Institute for Advanced Study, UCAS, Hangzhou 310024, China}
\affiliation{Institute of Cosmology and Gravitation, University of Portsmouth,  Portsmouth PO1 3FX, United Kingdom}
\affiliation{International Centre for Theoretical Physics Asia-Pacific, Beijing/Hangzhou, China}

\begin{abstract}
\noindent The symmetron is a light scalar which provides a screening mechanism so as to evade the strong constraints from local gravity tests. In order to achieve this goal, a $Z_2$ symmetry is imposed on the symmetron model. In this paper, we introduce a new symmetron Chern-Simons-like gravitational interaction which is $Z_2$ invariant but breaks the parity symmetry explicitly. As a result, it is found that this coupling can generate gravitational wave (GW) amplitude birefringence when GWs propagate over the symmetron backgrounds. Due to the matter density difference, the symmetron profile changes significantly when entering the galaxy, so that we need to discuss the extra-galactic and galactic situations separately. On the one hand, the cosmological symmetron field follows the adiabatic solution, which induces a parity-violating GW amplitude correction with its exponent proportional to the GW frequency and the traveling distance. On the other hand, the symmetron takes the screening solution within the Milky Way, and the generated GW birefringence is only a function of the GW frequency. By further comparing these two contributions, we find that the extra-galactic symmetron field produces the dominant birefringence effects. Finally, with the latest GW data from LIGO-Virgo-Kagra, we place a reasonable constraint on the parity-violating coupling parameter in this symmetron model. 



\end{abstract}
\maketitle

\newpage



\section{Introduction}\label{sec1}
The observation of gravitational waves (GWs) by the LIGO-Virgo-Kagra Collaboration (LVK) has opened a new avenue to test the nature of gravity\cite{LIGOScientific:2016aoc,LIGOScientific:2017zic,LIGOScientific:2018mvr,LIGOScientific:2020ibl,KAGRA:2021vkt}. One interesting new physics beyond General Relativity (GR) is the parity violation in gravity, which has been predicted in various modified gravity theories~\cite{Nojiri:2017ncd}, such as the Chern-Simons (CS) gravity~\cite{Lue:1998mq,Jackiw:2003pm,Alexander:2009tp}, Teleparallel Gravity~\cite{Conroy:2019ibo,Crisostomi:2017ugk}, ghost-free scalar-tensor gravity~\cite{Nishizawa:2018srh,Zhao:2019xmm,Qiao:2021fwi,Crisostomi:2017ugk} and many others~\cite{Horava:2009uw,Takahashi:2022mew,Bombacigno:2022naf,Boudet:2022nub,Kawai:2017kqt,Kostelecky:2016kfm,Altschul:2009ae,Sulantay:2022sag,Hojman:1980kv,Takahashi:2009wc,Zhu:2013fja}. Especially, these theories usually predict a remarkable phenomenon called GW birefringence~\cite{Alexander:2004wk,Alexander:2007kv,Yunes:2010yf,Alexander:2017jmt,Ezquiaga:2021ler}, {\it i.e.}, the two circular GW polarizations evolve differently in phase and amplitude when they propagate over astrophysical distances. Recently, such an effect has also been studied in Refs.~\cite{Satoh:2007gn,Kato:2015bye,Yoshida:2017cjl,Jung:2020aem,Chu:2020iil,Li:2020xjt,Ng:2023jjt,Wang:2020cub,Jenks:2023pmk,Callister:2023tws,Okounkova:2021xjv,Nojiri:2019nar,Daniel:2024lev,Manton:2024hyc,Lagos:2024boe,Huang:2024lzu} .  


Light scalars are another kind of predictions in many popular modified gravity theories (see {\it e.g.}, Refs.~\cite{Jain:2010ka,Clifton:2011jh,Joyce:2014kja,Koyama:2015vza,Brax:2021wcv,Baker:2019gxo} for recent reviews), like the superstring and supergravity models~\cite{Green:1987mn,Polchinski:1998rr}. They can play very important roles in the early and late evolutions of the Universe~\cite{Clifton:2011jh,Joyce:2014kja,Koyama:2015vza,Copeland:2006wr}. However, one problem plaguing these theories is that they typically leads to long-range scalar forces between matters, whose strengths are similar in magnitude to the Newtonian force so that they have been strongly constrained by local gravity tests, such as the fifth-force experiments~\cite{Will:2014kxa}. In the literature, there have been many nonlinear attempts to solve this problem (see {\it e.g.}, Refs.~\cite{Joyce:2014kja,Koyama:2015vza,Brax:2021wcv} for recent reviews and references therein), including the Vainshtein mechanism~\cite{Vainshtein:1972sx,Deffayet:2001uk,Arkani-Hamed:2002bjr}, the chameleon mechanism~\cite{Khoury:2003aq,Khoury:2003rn,Mota:2006ed,Burrage:2017qrf}, the environment-dependent dilaton~\cite{Damour:1994zq,Gasperini:2001pc,Damour:2002mi,Damour:2002nv,Brax:2010gi,Brax:2011ja,Brax:2022uyh}, the Galileon field~\cite{Dvali:2000hr,Nicolis:2008in,Ali:2012cv} and so on. 

More recently, the symmetron~\cite{Khoury:2003aq,Khoury:2003rn} provides a new idea to conceal itself from the local gravity detection (see {\it e.g.} Ref.~\cite{Burrage:2017qrf} for a recent review and references therein). By implementing a $Z_2$ symmetry acting on the symmetron, the coupling of its induced scalar force is proportional to the symmetron vacuum expectation value (VEV). It turns out that, at regions with low matter densities, the $Z_2$ invariance is spontaneously broken with a nonzero VEV. Thus, the scalar force becomes comparable to the Newtonian one, so that the symmetron can leave imprints in the cosmological and galactic dynamics~\cite{Burrage:2023eol,Burrage:2018zuj,Brax:2012nk,Davis:2011pj}. On the other hand, when the matter density increases to exceed a critical value, the symmetron VEV vanishes so that the $Z_2$ symmetry is recovered. In this case, the symmetron-induced force is severely suppressed, which could help to be consistent with the experimental data in the Solar system and in the Milky Way (MW).  

Inspired by the significance of the $Z_2$ symmetry in the symmetron model, it is quite natural to introduce a new $Z_2$-invariant CS-like coupling $\sigma^2 \tilde{R} R$~\cite{Burrage:2017qrf}, in which $\sigma$ denotes the symmetron while $\tilde{R}R$ refers to the Pontryagin density. Note that this coupling breaks the parity invariance explicitly. This is in contrast with the conventional CS interaction involving a pseudoscalar axion-like particle, which is parity symmetric but only violates this symmetry spontaneously in the presence of nontrivial pseudoscalar backgrounds~\cite{Nojiri:2019nar,Alexander:2007kv,Alexander:2017jmt,Yunes:2010yf}. Even though the origin of parity violation is different, the symmetron is still expected to generate the GW birefringence effect, which is the main focus on the present paper. Especially, as shown in Ref.~\cite{Hinterbichler:2011ca,Wang:2012kj}, the symmetron field would take highly nontrivial configurations either in the vast cosmological space or in our Galaxy, both of which could affect the propagation of GW signals observed at the Earth. Therefore, we would like to investigate novel characteristics of the GW birefringence induced by these symmetron profiles. Finally, given the precise measurements performed by the LVK Collaboration~\cite{LIGOScientific:2018mvr,LIGOScientific:2020ibl,KAGRA:2021vkt}, we also hope to utilize the existing GW data to constrain the introduced parity-violating coupling.

This paper is structured as follows. In Sec.~\ref{secAction} we shall give the symmetron action. Especially, we will introduce a new parity-violating $Z_2$-invariant gravitational coupling involving the symmetron. 
Sec.~\ref{SecSymm} is dedicated to briefly reviewing the symmetron profiles inside and outside of the MW, which serves as backgrounds for the GW propagation. Then we shall compute the GW birefringence under the influence of the galactic and extra-galactic symmetron fields, and use the latest data from LVK observations to constrain the model parameters in Sec.~\ref{SecGWB}. Finally, we conclude in Sec.~\ref{secCon}.

\section{Symmetron and its Gravitational Chern-Simons Coupling}\label{secAction}
Light scalars are ubiquitous in many extensions of GR, such as string theory. A nice way to hide such light scalars from the precise detection in the local Solar system is provided by the symmetron mechanism. In this mechanism, the light symmetron field $\sigma$ is conventionally implemented by a ${Z}_2$ symmetry: $\sigma\to -\sigma$, which dictates the following scalar-tensor action~\cite{Hinterbichler:2010es,Hinterbichler:2011ca,Burrage:2017qrf}:
\begin{eqnarray}
	S_0 &=& \int d^4 x \sqrt{-g} \left[\kappa R -\frac{1}{2} g^{\mu\nu} \partial_\mu \sigma \partial_\nu \sigma - V(\sigma) \right] 
	 + \int d^4 x \sqrt{-\tilde{g}} \mathcal{L}_{\rm m} (\psi, \tilde{g}_{\mu\nu})\,
\end{eqnarray} 
where $g_{\mu\nu}$ denotes the metric tensor in the Einstein frame, and $\kappa \equiv (16\pi G)^{-1} = M^2_{\rm Pl}/2$ with $G$ and $M_{\rm Pl}$ as the Newtonian constant and the reduced Planck mass, respectively. As detailed below, the symmetron mechanism is achieved by implementing the following ${Z}_2$ symmetric potential $V(\sigma)$~\cite{Hinterbichler:2011ca}
\begin{eqnarray}
	V(\sigma) = -\frac{1}{2}\mu^2 \sigma^2 + \frac{\lambda}{4} \sigma^4 +\frac{\mu^4}{4\lambda} = \frac{\lambda}{4}\left(\sigma^2 - \frac{\mu^2}{\lambda}\right)^2\,,
\end{eqnarray}
where $\mu$ represents a mass scale while $\lambda$ is a dimensionless coupling. Moreover, in order to guarantee the weak equivalence principle, the matter fields collectively written as $\psi$ should couple universally to the Jordan-frame metric $\tilde{g}_{\mu\nu} \equiv A^2 (\sigma) g_{\mu\nu}$~\cite{Postma:2014vaa} where the coupling function $A(\sigma)$ is taken in the following quadratic form~\cite{Khoury:2003aq,Khoury:2003rn,Mota:2006ed}
\begin{eqnarray}\label{}
	A(\sigma) = 1 + \frac{\sigma^2}{2M^2}\,,
\end{eqnarray}  
with $M$ as the UV cutoff scale of the theory. As mentioned in Ref.~\cite{Burrage:2017qrf}, due to this ${Z}_2$ symmetry, it is quite natural to introduce the following CS-like coupling:
\begin{eqnarray}\label{CS-action}
	S_{\rm CS} =  \int d x^{4} ~ \sqrt{-g} ({\alpha}/{4})\sigma^{2}{R}^{\tau}_{~\lambda \mu\nu}\widetilde{R}^{\lambda~\mu\nu}_{~\tau}\, ,
\end{eqnarray} 
where $\alpha$ denotes the coupling constant. Here $\widetilde{R}^{\lambda}{}_{\tau}{}^{\mu\nu} \equiv \epsilon^{\mu\nu\rho\sigma}R^{\lambda}_{~\tau\rho\sigma}/2$ is the dual Riemann tensor and ${R}^{\tau}_{~\lambda \mu\nu}\widetilde{R}^{\lambda~\mu\nu}_{~\tau}$ is the Pontryagin density,  where $\epsilon^\mathrm{\mu\nu\rho\sigma}=\tilde{\epsilon}^{\mu\nu\rho\sigma}/\sqrt{-g}$ is the 4-dimensional Levi-Civita tensor with the anti-symmetric symbol defined as $\tilde{\epsilon}^{0123}=-\tilde{\epsilon}_{0123}=1$. It is worth mentioning that this coupling breaks the parity symmetry explicitly. In contrast, the conventional gravitational CS interaction~\cite{Alexander:2009tp} with a single axion-like pseudoscalar is parity symmetric, and parity is spontaneously broken only in the presence of a nontrivial pseudoscalar background. In sum, the total symmetron action can be written as 
\begin{eqnarray}\label{TotalAction}
	S = S_0+ S_{\rm CS}\,.
\end{eqnarray}

\section{Symmetron Fields Inside and Outside the Milky Way}\label{SecSymm}
Before delving into the GW birefringence and related parity-violating phenomena, we need firstly specify the symmetron backgrounds for the GW propagation. 
By varying Eq.~(\ref{TotalAction}) with respect to $\sigma$, we can obtain the following equation of motion: 
\begin{align}\label{EoMs}
	\square\sigma=\frac{\partial V}{\partial \sigma}-A^3(\sigma)\frac{\partial A}{\partial\sigma}\tilde{T} - \frac{\alpha}{\kappa} \sigma \tilde\epsilon^{\mu\nu\rho\xi} R^\tau_{\ \lambda\mu\nu}R^\lambda_{\ \tau\rho\xi}\,,
\end{align}
where $\tilde{T}\equiv\tilde{g}^{\mu\nu}\tilde{T}_{\mu\nu}$ is the trace of the stress-energy tensor $\tilde{T}_{\mu\nu}\equiv-(2/\sqrt{-\tilde{g}})\delta\mathcal{L}_{\mathrm{m}}/\delta\tilde{g}^{\mu\nu}$ in the Jordan frame. 
For each energy component $i$, this quantity is given by $\tilde{T}_i = (-1+3w_i)\tilde{\rho}_i$ with $w_i$ as the corresponding equation of state (EoS). As a result, the solution to the conservation equation $\tilde{\nabla}_{\mu}\tilde{T}_{i\,\nu}^{\mu}=0$ is given by $\tilde{\rho} \propto A^{-3(1+w_i)} a^{-3(1+w_i)}$. Therefore, we can define $\rho_i \equiv A^{3(1+w_i)}\tilde{\rho}_i$~\cite{Hinterbichler:2011ca,Hui:2009kc,Sakstein:2014isa}, so that $\rho_i$ follows the conventional evolution $\rho \propto a^{-3(1+w_i)}$ as in the standard cosmology. 
Meanwhile, due to the suppression by the Planck mass in $\alpha/\kappa$, the last term in Eq.~(\ref{EoMs}) is expected to be so small that we can safely neglect its role in determining the symmetron configurations. As a result, the symmetron equation can be simplified as follows
\begin{eqnarray}\label{EqSym0}
	\square \sigma = \frac{\partial V}{\partial \sigma} + \sum_i \frac{1}{A^{3w_i}} \frac{\partial A}{\partial \sigma} (1- 3w_i) \rho_i \,.
\end{eqnarray}
Finally, caused by the screening mechanism in the high-density region, the symmetron field behaves quite differently outside and inside the MW, so that we would like to study these two cases separately. This part of discussions is not new, but serves as the background for our following investigation of the GW birefringence over the symmetron cosmology.


\subsection{Cosmological Symmetron Profile}\label{SecSymmCos}
Outside of the MW, the symmetron is expected to be unscreened. Over cosmological scales, the spacetime and the symmetron profile are assumed to be homogeneous and isotropic. As demonstrated in Eq.~\eqref{EoMs}, the evolution of the symmetron $\sigma$ is controlled by the following effective potential~\cite{Hinterbichler:2011ca}
\begin{eqnarray}\label{Veff}
	V_{\rm eff} (\sigma) = V({\sigma}) + \sum_i A^{1-3w_i} (\sigma) \rho_i \,,
\end{eqnarray}
where the summation is over the cosmological energy density contents $i$.  Since our focus here is the late Universe with the redshift $z\lesssim 1$, the total energy is dominated by the dark and visible matter $\rho_m$ with its EoS $w_m = 0$ as well as the dark energy $\rho_{\rm de}$ which drives the present cosmic acceleration. Here we assume that the dark energy consist of the cosmological constant with $w_{\rm de} = -1$. Based on the Friedmann equation, we can represent the dark energy density as $\rho_{\rm de} = \rho_c (1-\Omega_m)$, where $\rho_c \equiv 3M_{\rm Pl}^2 H_0^2$ is the critical density for a flat Universe and $\Omega_m$ denotes the current fractional density of matter. 
Moreover, it is shown in Refs.~\cite{Hinterbichler:2011ca,Wang:2012kj} that after the Universe comes into the dark energy dominated era, the symmetron would soon be saturated into the following adiabatic solution,
\begin{eqnarray}\label{SolAd}
	\sigma^2 = \frac{1}{\lambda} \left( \mu^2 - \frac{\rho_c [\Omega_m a^{-3}+4 (1-\Omega_m)]}{M^2}  \right)\,,
\end{eqnarray}
where we have approximated the coupling function $A(\sigma) \approx 1$ for $\sigma_0 \ll M$. This solution tracks the minimum of the effective potential in Eq.~\eqref{Veff}, and is the attracting point of the  symmetron dynamics~\cite{Hinterbichler:2011ca,Wang:2012kj}. In the following, we shall make use of this adiabatic solution as the background for the extragalactic propagation of GWs.

\subsection{Symmetron Profile Around the Milky Way}\label{SecSymmMW}
Around astrophysical objects, like the Solar system or the MW, the matter density is so high that the symmetron screening mechanism becomes effective. Here we assume the distribution is spherically symmetric, so that the symmetron field should also follow this property. 
Note that the scalar profile here can be viewed as static, since the evolution time scale of our galaxy is much larger than that for the GW galactic propagation. Moreover, over the galactic scale, the cosmological expansion can be ignored, and the background metric can be taken to be of Minkowskian form, ${\it e.g.}$, $g_{\mu\nu} = \eta_{\mu\nu} = {\rm diag}(-1,1,1,1)$. As a result, the equation governing the symmetron profile in Eq.~\eqref{EqSym0} is simplified as follows~\cite{Hinterbichler:2010es}
\begin{eqnarray}\label{EqSymm}
	\frac{d^2 \sigma}{dr^2} + \frac{2}{r}\frac{d\sigma}{dr} - \frac{dV_{\rm eff}}{d\sigma} = 0\,, 
\end{eqnarray} 
where the effective potential is defined as $V_{\rm eff} = A(\sigma) \rho_m + V (\sigma)$ as the local density $\rho_m$ is dominated by the matter mass.
Inside the galaxy where its mass density is assumed to be homogeneous, the effective potential can be approximated to be of quadratic form
\begin{eqnarray}\label{Vin}
	V_{\rm eff} (\sigma) =\left(\frac{\rho_m}{M^2} - \mu^2\right)\sigma^2\,,
\end{eqnarray} 
while outside of the MW the matter density nearly vanishes so that the effective potential near the vacuum can be written as
\begin{eqnarray}\label{Vout}
	V_{\rm eff} (\sigma) = \mu^2 (\sigma-\sigma_0)^2\,,
\end{eqnarray}
in which the vacuum expectation value of the symmetron is
\begin{eqnarray}\label{sig0}
	\sigma_0 = \mu/\sqrt{\lambda}\,.
\end{eqnarray}
Furthermore, the appropriate boundary conditions are given by~\cite{Hinterbichler:2011ca}
\begin{eqnarray}\label{Cbdy}
	\left.\frac{d\sigma}{dr}\right|_{r=0} = 0\,,\quad\quad \sigma(r\to \infty) = \sigma_0\,.
\end{eqnarray}
Consequently, by solving the differential equation in Eq.~\eqref{EqSymm} with effective potentials in Eqs.~\eqref{Vin} and \eqref{Vout} and boundary conditions in Eq.~\eqref{Cbdy}, the symmetron profile inside the MW is given by~\cite{Hinterbichler:2011ca}
\begin{eqnarray}\label{Sin}
	\sigma_{\rm in} = C_{\rm in} \frac{R}{r} \sinh\left( r\sqrt{\frac{\rho_m}{M^2}-\mu^2} \right)\,,
\end{eqnarray}
while the scalar configuration external to the galaxy is
\begin{eqnarray}\label{Sout}
	\sigma_{\rm out} = \sigma_0 + C_{\rm out} \frac{R}{r} e^{-m_0 (r-R)}\,.
\end{eqnarray}
By matching these two solutions at the galactic radius $r= R$, we can determine the parameters in these expressions
\begin{eqnarray}\label{CIO}
	C_{\rm in} &=& \frac{\sigma_0}{\cosh(\sqrt{R/\Delta R})} \sqrt{\frac{\Delta R}{R}}\,, \nonumber\\
	C_{\rm out} &=& - \sigma_0 \left[1- \sqrt{\frac{\Delta R}{R}} \tanh\left( \sqrt{\frac{R}{\Delta R}} \right) \right]\,.
\end{eqnarray}
where 
\begin{eqnarray}\label{ThinShell}
	\frac{\Delta R}{R} \equiv \frac{M^2}{\rho_m R^2} = \frac{M^2}{6 M_{\rm Pl}^2 \Phi_N} \,.
\end{eqnarray} 
For a strongly gravitating object, the typical mass density and the associated surface Newtonian potential $\Phi_N$ are so large that the ratio $\Delta R/R \ll 1$. In this case, the scalar force induced by the symmetron is screened, which can be proved by considering the ratio of the symmetron-induced scalar force with the Newtonian one acting on a test particle at the distance $R \ll r \ll m_0^{-1} $ with $m_0 = \sqrt{2}\mu$ denoting the symmetron mass
\begin{eqnarray}
	\frac{F_\sigma}{F_N} &=& \frac{-\nabla \ln A(\sigma)}{-\nabla \Phi_N} = \frac{\sigma (d\sigma/dr)/AM^2}{d\Phi_N /dr} \approx -\frac{6M_{\rm Pl}^2}{M^2} \frac{C_{\rm out}\sigma_0}{R^2 \rho_m} \nonumber\\
	&=& 6g^2 \frac{\Delta R}{R} \left[ 1- \sqrt{\frac{\Delta R}{R}} \tanh\left( \sqrt{\frac{R}{\Delta R}} \right) \right] \ll 1\,.
\end{eqnarray}
where the Newtonian potential is given by
\begin{eqnarray}
	\Phi_N =- \frac{(4\pi/3)G_N R^3 A(\sigma) \rho_m}{r}\,,
\end{eqnarray}
while the scalar outside of the object is approximated as
\begin{eqnarray}
	\sigma = - \frac{\sigma_0 R}{r} e^{-m_0 r} \approx -\frac{\sigma_0 R }{r}  \,.
\end{eqnarray}
The extremely small ratio of $F_\sigma/F_N$ indicates that the scalar force is screened in the high-density region. However, for low-density objects where $\Delta R/R \gg 1$, the scalar field profile external to the object is given by
\begin{eqnarray}
	\sigma_{\rm out} = \sigma_0 \left(1 - \frac{1}{3}\frac{R}{\Delta R} \frac{R}{r} e^{-m_0 (r-R)}  \right)\,.
\end{eqnarray} 
In this case, the ratio $F_\sigma/F_N$ can be approximated as
\begin{eqnarray}
	\frac{F_\sigma}{F_N} = 2 M_{\rm Pl}^2 \frac{\sigma_0^2}{M^4} = 2 g^2\,.
\end{eqnarray}
Here we assume that there should be no screening for small-density regions, so that the scalar force should be comparable to the Newtonian one with $F_\sigma/F_N \approx 1$,  which implies $g\sim {\cal O}(1)$. 

We further require the transition from the early screened Universe to the late unscreened one at the recent epoch with the redshift $z\approx 1$, which gives the following relation
\begin{eqnarray}\label{RadTrans}
	\mu^2 \sim \frac{\rho_c}{M^2} \sim \frac{H_0^2 M_{\rm Pl}^2}{M^2}\,.
\end{eqnarray}

Now we can estimate the orders of parameters appearing in this model. From the local gravity tests~\cite{Will:2014kxa,Bertotti:2003rm}, we can obtain the following constraint
\begin{eqnarray}\label{ConsR}
	\frac{\Delta R}{R} \lesssim 6\times 10^{-3}\,.
\end{eqnarray}  
By using the fact that the surface gravity in the MW is $\Phi_N \sim {\cal O}(10^{-6})$~\cite{Hinterbichler:2011ca}, the relation in Eq.~\eqref{ThinShell} gives the following upper bound on the cutoff scale
\begin{eqnarray}\label{ConsM}
	\frac{M}{M_{\rm Pl}} \lesssim 1.9 \times 10^{-4}  \,.
\end{eqnarray}
Thus, the symmetron mass in the vacuum can be estimated as
\begin{eqnarray}\label{mSymm}
	m_0 = \sqrt{2}\mu \gtrsim 10^3 H_0 \gtrsim 1~{\rm Mpc}^{-1}\,,
\end{eqnarray}
which indicates that the symmetron Compton wavelength should be smaller than Mpc scale. Notice that this result was further confirmed by applying another argument in Ref.~\cite{Wang:2012kj}.

\section{Gravitational Wave Birefringence}\label{SecGWB}
Now we turn to the GW birefringence in the presence of the symmetron backgrounds inside and outside of our galaxy. By implementing a new $Z_2$-symmetric but parity-violating gravitational CS coupling in Eq.~\eqref{CS-action}, we expect that it would give rise to the GW birefringence over the symmetron fields. By differentiating the total action in Eq.~\eqref{TotalAction} with respect to the tensor perturbation $h_{ij}$ defined as $ds^2=a(\eta)^2[-d\eta^2+(\delta_{ij}+h_{ij})]$, the GW equations of motion can be obtained as follows~\cite{Jenks:2023pmk}
\begin{align}
	\label{EoMGWt}
	\square h^j_{~i} =  \frac{\alpha}{\kappa} \epsilon^{pjk} \partial^\alpha \left[\frac{1}{a^2} \left( \partial_p \sigma^2 \partial_\eta \partial_\alpha - \partial_\eta \sigma^2 \partial_p  \partial_\alpha \right) \right]h_{ki}\,,
\end{align}
where $\square \equiv (-\partial^{2}_{\eta} - 2\mathcal{H}\partial_{\eta} + \partial^{2}_{i})/a^2$ 
is the four-dimensional d’Alembertian operator of the Freedman-Lema\^{i}tre-Robertson-Walker metric. For convenience, we choose the following transverse-traceless gauge conditions, 
\begin{align}
	\delta^{ij}h_{ij}=0\,,\quad \partial^{i}h_{ij}=0\,.
\end{align}
We also set the coordinate system in which the GW propagates along the $z$ direction 
so that the physical plus- and cross-polarizations can be given by
\begin{align}\label{GWform1}
	h_{\mu\nu} = \left(\begin{matrix}
		0 & 0          & 0           & 0 \\
		0 &    h_{+}   &  h_{\times} & 0 \\
		0 & h_{\times} &    -h_{+}   & 0 \\
		0 &     0      &     0       & 0
	\end{matrix}\right)e^{-i(\omega \eta - k z)}\,.
\end{align}
Moreover, it is more useful to combine these linearized polarizations into the right- and left-handed circularly polarized states as~\cite{Isi:2022mbx}
\begin{align}\label{GWformLR}
	h_{\rm R,L} = \frac{1}{\sqrt{2}} (h_{+} \mp  i h_{\times})\,.
\end{align}
It turns out that the GW equation of motion in Eq.~\eqref{EoMGWt} can be reorganized into two separate equations for two circular polarizations as follows
\begin{align}\label{EoMLR}
	\square h_{\rm R,L} =   i\lambda_{\rm R,L} \frac{\alpha}{\kappa}\partial^\alpha \left[\frac{1}{a^2} \left( \partial_p \sigma^2 \partial_\eta \partial_\alpha - \partial_\eta \sigma^2 \partial_p  \partial_\alpha \right) \right]h_{\rm R,L} \, ,
\end{align}
in which $\lambda_{\rm R,L} = \pm 1$. Further simplifications can be achieved by specifying the symmetron scalar backgrounds inside and outside the MW, both of which have already been detailed in Sec.~\ref{SecSymm}. We shall consider these two situations separately in what follows.

\subsection{GW Birefringence Outsides of the Milky Way}\label{GWextra}
In this subsection, we will investigate the parity-violating birefringence of GWs when they move over the extra-galactic symmetron background. Since GWs would propagate over cosmological distances, one can take the geometry and the scalar profile to satisfy the cosmological principle, {\it i.e.}, both are homogeneous and isotropic on average over large scales. In other words, the symmetron $\sigma$ should be only the function of the conformal time $\eta$. Thus, the propagating equations for GWs of both handednesses in Eq.~\eqref{EoMLR} can be simplified into the following form
\begin{eqnarray}
	\square h_{\rm R,L} = - \frac{i\lambda_{\rm R,L}\alpha}{\kappa a^2} \left[-\frac{1}{a^2} \left[(\sigma^2)^{\prime\prime} - 2 {\cal H}(\sigma^2)^\prime\right] \partial_z h^\prime_{\rm R,L} + (\sigma^2)^\prime \square \partial_z h_{\rm R,L} \right]\,,
\end{eqnarray}
where the prime denotes the derivative with respect to $\eta$. When putting the GW waveforms of Eqs.~\eqref{GWform1} and \eqref{GWformLR} into the above equation,  we can yield the modified GW dispersion relations 
\begin{eqnarray}\label{Dispersion0}
	\left[1- \lambda_{\rm R,L} (\alpha/\kappa a^2) (\sigma^2)^\prime k\right] (\omega^2 + 2i\mathcal{H}\omega -k^2) = i\lambda_{\rm R,L} \alpha \left[(\sigma^2)^{\prime\prime} - 2\mathcal{H}(\sigma^2)^\prime\right] \omega k/(\kappa a^2)\,.
\end{eqnarray}
As argued in Sec.~\ref{SecSymmCos}, the symmetron in the cosmology should follow the adiabatic evolution in Eq.~\eqref{SolAd}, so that we have 
\begin{eqnarray}
	(\sigma^2)^\prime = \frac{3\rho_c \Omega_m}{\lambda M^2} \frac{\cal H}{a^3}\,, \quad 	\left(\sigma^2\right)^{\prime\prime} = \frac{3\rho_c \Omega_m}{\lambda M^2 a^3} \left( \mathcal{H}^\prime - {3\mathcal{H}^2}\right)\,.
\end{eqnarray}
Since the present LVK GW events are all around 1 Gpc far away from the Earth, we can take the small redshift limit~\cite{Jenks:2023pmk} where the Hubble parameter $H=\mathcal{H}/a$ is taken to be constant. Thus, $\mathcal{H}^\prime$ can approximated as follows
\begin{eqnarray}
	\mathcal{H}^\prime = \frac{d\mathcal{H}}{d\eta} = a^\prime \frac{d\mathcal{H}}{da} \approx a^\prime H = {\mathcal{H}^2}\,.
\end{eqnarray} 
By taking all of these approximations into Eq.~\eqref{Dispersion0}, we can obtain the following relation
\begin{eqnarray}\label{Dispersion1}
	\omega  &\approx&  k - \frac{i}{2} \left\{2\mathcal{H} - \lambda_{\rm R,L} \left(\frac{\alpha \omega}{\kappa a^2}\right) \left[ (\sigma^2)^{\prime\prime} - 2\mathcal{H}(\sigma^2)^\prime \right] \right\} \nonumber\\
	& \approx & k-  i\mathcal{H} + \frac{i\lambda_{\rm R,L}}{2} \left(\frac{12 \alpha \omega \rho_c \Omega_m \mathcal{H}^2 }{\kappa \lambda M^2 a^5}\right)  \equiv k - i\mathcal{H} + \Delta \omega\,.
\end{eqnarray}
which is valid up to the leading order in the coupling $\alpha/\kappa$. Thus, the extra-galactic symmetron background only generates the following imaginary variations to the GW phase
\begin{eqnarray}\label{DSc}
	\Delta S_c &=& -\int^{\eta_0}_{\eta_e} \Delta \omega d\eta  = i \lambda_{\rm R,L}\left(\frac{12\alpha\omega \rho_c \Omega_m}{2\kappa \lambda M^2}\right) \int^{\eta_0}_{\eta_e} d\eta \frac{\mathcal{H}^2}{a^5} 
	\approx  i\lambda_{\rm R,L} \left(\frac{12\alpha\omega \rho_c \Omega_m H_0^2 d_c}{2\kappa \lambda M^2}\right) \,,
\end{eqnarray} 
where we have used the following relations
\begin{eqnarray}
	\int^{\eta_0}_{\eta_e} d\eta \frac{\mathcal{H}^2}{a^n} = \int^{a_0}_{a_e} \frac{\mathcal{H}da}{a^{n+1}} \approx H_0 \int^{a_0}_{a_e}\frac{da}{a^{n}} =  \frac{H_0}{n-1} \left(\frac{1}{a_e^{n-1}} - \frac{1}{a_0^{n-1}}\right) \approx H_0 z  \approx H_0^2 d_c  
\end{eqnarray}
in the small redshift limit. Therefore, the left- and right-handed polarized GW waveforms are changed by
\begin{eqnarray}\label{GWcos}
	h_{\rm R,L} (f) = h_{\rm R,L}^{\rm GR} e^{i\Delta S_c} \approx h_{\rm R,L}^{\rm GR} \exp\left(-\lambda_{\rm R,L} \kappa_A \times \frac{f}{\rm 100~Hz} \times \frac{d_c}{\rm 1~Gpc} \right) \,,
\end{eqnarray}
where $f = \omega/(2\pi)$ is the GW frequency and we have defined the opacity parameter $\kappa_A$ as
\begin{eqnarray}\label{kaCos}
	\kappa_A \equiv \frac{12\pi \alpha}{\kappa} \frac{\rho_c \Omega_m H_0^2}{\lambda M^2} = \frac{72\pi \alpha\Omega_m H_0^4 }{\lambda M^2} \,.
\end{eqnarray}
It is worth noting that the phase variation in Eq.~\eqref{DSc} is imaginary, which means that the symmetron with its new CS-like coupling in Eq.~\eqref{CS-action} does not induce any velocity birefringence. Rather, it leads to the GW amplitude birefringence, with the final birefringence factor in Eq.~\eqref{GWcos} similar to the conventional CS gravity for axion-like particles~\cite{Alexander:2009tp,Jenks:2023pmk,Alexander:2004wk,Alexander:2007kv,Wang:2020cub}.  

\subsection{GW Birefringence Inside the Milky Way}\label{GWin}
Besides of the extra-galactic contribution to the GW birefringence, the symmetron profile around the MW would give rise to the additional effects to GWs. It has been shown in Sec.~\ref{SecSymmMW} that, caused by the screening mechanism, the symmetron profile would be dramatically altered near the boundary of our Galaxy, with the explicit solutions given in Eqs.~\eqref{Sin}, \eqref{Sout} and \eqref{CIO}. Here we can approximate the solution to be static, since the GW propagation time inside the MW is much shorter than the typical time scale of the galaxy evolution. For the same reason, the cosmic expansion can be ignored at the galactic scale, so that the metric can be taken to be flat. Therefore, we can reduce the general equations of motion for GWs in Eq.~\eqref{EoMLR} into the following form
\begin{align}
	\square h_{\rm R,L} -  i \lambda_{\rm R,L} (2\alpha/\kappa) \partial^\alpha ( \sigma \partial_z \sigma \partial_\alpha \partial_t  h_{\rm R,L} )= 0\,.
\end{align}
where we have set the scale factor $a=1$ and replaced the conformal time $\eta$ with the physical one $t$ so that the d'Alembertian operator is given by $\square \equiv -\partial_{t}^{2} + \partial_{i}^{2}$. As a consequence, the GW dispersion relation is modified according to 
\begin{eqnarray}
	\omega^2 - k^2 \approx i\lambda_{\rm R,L} ({\alpha}/{\kappa}) \omega k \left[\partial_z^2 \sigma^2\right]\,,
\end{eqnarray}
which further leads to
\begin{eqnarray}
	k  \approx \omega - i\lambda_{\rm R,L} (\alpha\omega/2\kappa) \left[\partial_z^2 \sigma^2\right] \equiv \omega + \Delta k\,.
\end{eqnarray}
The GW phase variation within the Galaxy is given by
\begin{eqnarray}
	\Delta S_g = \int^{{\bf x}_{\rm in}}_{{\bf x}_{\rm out}} \Delta k_i dx^i = -i\lambda_{\rm R,L} (\alpha\omega/\kappa) (\sigma_{\rm in} \partial_z \sigma_{\rm in} - \sigma_{\rm out} \partial_z \sigma_{\rm out})\,,
\end{eqnarray} 
and the GWs of both circular polarizations are modified to
\begin{eqnarray}\label{GWgal}
	h_{\rm R,L}(f) = h^{\rm GR}_{\rm R,L} (f) e^{i\Delta S_g} \approx h_{\rm R,L}^{\rm GR} (f) \exp\left( \lambda_{\rm R,L} \kappa_A^\prime \frac{f}{\rm 100 Hz} \right)\,, 
\end{eqnarray}
where the associated opacity parameter $\kappa_A^\prime$ is defined as
\begin{eqnarray}\label{kapP}
	\kappa_A^\prime \equiv (2\pi\alpha/\kappa) (\sigma_{\rm in} \partial_z \sigma_{\rm in} - \sigma_{\rm out} \partial_z \sigma_{\rm out})\,.
\end{eqnarray} 
Here $\sigma_{\rm in,\,out}$ refers to the symmetron scalar values at the Solar system and far outside of the MW, which should take values of the screening solutions in Eqs.~\eqref{Sin} and \eqref{Sout}, respectively. On the one hand, this screening solution should smoothly connected to the cosmological scalar configuration studied in Sec.~\ref{SecSymmCos} which is homogeneous and isotropic. Thus, we should take the matching point sufficiently far away from the center of the galaxy so that the divergence of the scalar field becomes nearly zero, {\it i.e.}, $\partial_z \sigma_{\rm out} \simeq 0$, which leads the second term in Eq.~\eqref{kapP} to vanish identically. In this case, only the scalar field $\sigma_{\rm in}(r_{\odot})$ at the Solar position in Eq.~\eqref{Sin} contributes to Eq.~\eqref{GWgal}, where $r_{\odot}=8.5$~kpc denotes the distance of the Sun from the galactic center. According to the discussion in Sec.~\ref{SecSymmMW}, we have the parameter hierarchy $\mu^2 \sim \rho_c/M^2 \ll \rho_m/M^2$ for the galactic matter density $\rho_m$, giving us the following approximated expression
\begin{eqnarray}
	\sigma_{\rm in} (r_\odot) \simeq \frac{\sigma_0}{\cosh(\sqrt{R/\Delta R})} \sqrt{\frac{\Delta R}{R}} \frac{R}{r_\odot} \sinh\left( \frac{r_\odot}{R} \sqrt{\frac{R}{\Delta R}} \right)\,,
\end{eqnarray}
where we have used the definition of $\Delta R/R$ in Eq.~\eqref{ThinShell}. Therefore, the opacity parameter $\kappa^\prime_A$ can be given by
\begin{eqnarray}
	\kappa^\prime_A =&& \frac{2\pi \alpha}{\kappa} \frac{\sigma_0^2}{r_\odot} \sqrt{\frac{\Delta R}{R}} \frac{R}{r_\odot} \frac{\sinh\left(\sqrt{\frac{R}{\Delta R}}\frac{r_\odot}{R}\right)}{\cosh^2\left(\sqrt{{R}/{\Delta R}}\right)}\left[ \cosh\left(\sqrt{\frac{R}{\Delta R}}\frac{r_\odot}{R}\right) - \sqrt{\frac{\Delta R}{R}} \frac{R}{r_\odot} \sinh\left(\sqrt{\frac{R}{\Delta R}}\frac{r_\odot}{R}\right)  \right] \nonumber\\
	&& \times \cos\left\langle {\bf k}, {\bf r}_\odot \right\rangle\,,
\end{eqnarray}
where the bracket $\langle {\bf k}, {\bf r}_\odot \rangle$ denotes the angle between the GW propagation direction ${\bf k}$ and the spatial divergence of the symmetron field $\mathbf{r}_\odot$ at the Solar system. Similar to the cosmological symmetron generated birefringence, the GWs propagating over the galactic symmetron configuration also experience the amplitude birefringence without any change in the velocity or in the phase of the gravitational waveform.

Note that the total correction to the GW amplitude is given by the multiplication of the birefringence factors in Eqs.~\eqref{GWcos} and \eqref{GWgal}, corresponding to the extra-galactic and galactic contributions, respectively. 
Let us now focus on the ratio of exponents in these two factors
\begin{eqnarray}
	\left|\frac{\kappa_A^\prime}{\kappa_A d_c}\right|  \sim \frac{1}{18} \frac{1}{\cosh^2\left(\sqrt{R/\Delta R}\right)}  \frac{1}{\Omega_m H_0^2 d_c r_{\odot}} \lesssim 10^{-5}\,,
\end{eqnarray}
where we have used the relations in Eqs.~\eqref{sig0}, \eqref{RadTrans}, \eqref{ConsR} and \eqref{ConsM} to simplify the expression. In this order-of-magnitude estimation, we have also taken the typical conformal distance from a black hole merger event to be of the LVK average value $d_c = 1.2$~Gpc~\cite{Lagos:2024boe}, while the present Hubble parameter is $H_0 = 0.2254\,{\rm Gpc}^{-1}$ with the matter density fraction $\Omega_m = 0.3$~\cite{ParticleDataGroup:2024cfk}. It is evident from this simple calculation that the GW amplitude modification for current LVK events are dominated by the extra-galactic contribution. This result can be easily understood by noticing that the birefringence factor in Eqs.~\eqref{GWgal} and \eqref{kapP} only depends on the local values of the symmetron field at the Solar system and far outsides of the MW, where the symmetron configurations already approach the $Z_2$-symmetric and $Z_2$-breaking values, respectively. Hence, the corresponding spatial variations $\partial_z \sigma_{\rm in, out}$ tend to vanish, resulting in significant suppression of the galactic contribution to the GW birefringence effect.



\subsection{Constraints on GW Birefringence}\label{CONstrain}
Currently there have already been stringent constraints on the GW birefringence in the literature~\cite{Okounkova:2021xjv,Ng:2023jjt,Lagos:2024boe} based on the LVK GW observations~\cite{LIGOScientific:2016aoc, LIGOScientific:2017zic, LIGOScientific:2018mvr, LIGOScientific:2020ibl, KAGRA:2021vkt} of compact binary coalescence events. Therefore, we expect that the same data would also constrain the parity-violating symmetron CS coupling of Eq.~\eqref{CS-action} introduced in this paper.

As mentioned previously in this section, the extra-galactic symmetron background gives the dominant contribution to the GW birefringence due to the accumulation of this effect over the cosmological distances. Moreover, it is seen from Eq.~\eqref{GWcos} that the exponent of the associated amplitude birefringence factor is proportional to the propagation distance $d_c$ and the frequency $f$ of GWs, which shares the same behavior as in the conventional CS gravity theories~\cite{Alexander:2009tp,Jenks:2023pmk,Alexander:2004wk,Alexander:2007kv,Wang:2020cub}. Thus,  we can apply most stringent constraints on the CS gravity in Refs.~\cite{Okounkova:2021xjv,Ng:2023jjt} directly to our present symmetron case. In particular, with the GWTC-2 Catalog, the authors in Ref.~\cite{Okounkova:2021xjv} employed the binary black hole merger data to give the upper bound on the opacity parameter as $\kappa_{A}\lesssim 0.74$. In a more recent study~\cite{Ng:2023jjt}, the GWTC-3 data was used to place a more stringent limit with $\kappa_{A}\lesssim 0.03$. Given the expression of $\kappa_A$ in Eq.~\eqref{kaCos} for the symmetron-induced birefringence, we can yield the following constraint on model parameters
\begin{eqnarray}
	\alpha (M/M_{\rm Pl})^4 \lesssim 8\times 10^{-22} ~{\rm Gpc}^2\,,
\end{eqnarray}
where we have used the result in \cite{Ng:2023jjt}. If we take the largest value of $M/M_{\rm Pl} \sim 1.9\times 10^{-4}$ implied by the constraint in Eq.~\eqref{ConsM} obtained by local gravity tests, the upper bound on the symmetron CS coupling in Eq.~\eqref{CS-action} is given by
\begin{eqnarray}\label{ConsPara}
	\alpha \lesssim 6\times 10^{-7}~ {\rm Gpc}^2\,.
\end{eqnarray}
If we further decrease the value of $M/M_{\rm Pl}$ as allowed by Eq.~\eqref{ConsM}, such a constraint on $\alpha$ can be even more relaxed.

\section{Conclusions}\label{secCon}
The symmeron provide us a viable screening mechanism that can conceal itself from the precise local gravity tests but still induce long-range forces which can leave imprint in the cosmological evolution. One important ingredient in this mechanism is the introduction of the $Z_2$ symmetry acting on the symmetron scalar. This motivates us to consider a new $Z_2$-invariant parity-violating gravitational coupling in Eq.~\eqref{CS-action}, which is expected to generate the GW birefringence when GWs move over symmetron backgrounds. Moreover, due to the screening of the scalar force, the symmetron profile would be changed dramatically around the MW, which indicates that we should divide our discussion into the extra-galactic and galactic parts. Outsides of the MW, the cosmological symmetron is believed to be homogeneous and isotropic, following the adiabatic evolution solution to the effective potential. It is found that such a symmetron background can induce the amplitude birefringence, with the exponent in the birefringence factor proportional to the GW frequency and the event distance. On the other hand, when GWs come into the MW, the highly nontrivial screening configuration gives rise to the additional contribution to the parity-violating amplitude correction, which is found to be only the function of GW frequency without any dependence on the propagating distance. By taking the ratio of these two birefringence factors, we have found that the cosmological symmetron produces the dominant effect on the GW amplitudes. Finally, with the most recent GW data from LVK observations, we can place useful constraints on the symmetron model parameters. In particular, by noticing that our symmetron model shares the same birefringence formula as in the conventional CS gravity, we can directly apply the existing constraints on the opacity parameter in Refs.~\cite{Okounkova:2021xjv,Ng:2023jjt} to our symmetron case, which can be translated into the upper limit on the new parity-violating coupling.

\begin{acknowledgments}
\noindent DH would like to thank Prof.~Kazuya Koyama for his warm hospitality at the Institute of Cosmology and Gravitation, University of Portsmouth, where part of this work was carried out. DH is also grateful to Prof. Ian Harry and Prof. Tessa Baker for enlightening discussions. This work is supported in part by the National Key Research and Development Program of China under Grant No. 2021YFC2203003, and the China Scholarship Council under Grant No. 202310740003.
\end{acknowledgments}

\appendix


\bibliography{GWsym}

\end{document}